\documentclass[aps,twocolumn,10pt,pre]{revtex4-1}

\usepackage{amsfonts} 
\usepackage{hyperref}
\usepackage{mdframed} 
\usepackage[percent]{overpic} 
\usepackage{xspace} 
\usepackage{amsmath}

\newcommand{\pcf}{\textsc{pcf}\xspace} 
\newcommand{\vpcf}{\textsc{vpcf}\xspace}

\begin{document}

\title{Pair correlation function based on Voronoi topology} 

\author{Vasco M. Worlitzer} 

\author{Gil Ariel} 

\author{Emanuel A. Lazar} 
\affiliation{Department of Mathematics, Bar Ilan University, Ramat Gan, Israel} 

\date{\today}

\begin{abstract} 
The pair correlation function (PCF) has proven an effective tool for analyzing 
many physical systems due to its simplicity and its applicability to simulated 
and experimental data. However, as an averaged quantity, the PCF can fail to 
capture subtle structural differences in particle arrangements, even when 
those differences can have a major impact on system properties. Here, we 
use Voronoi topology to introduce a discrete version of the PCF that 
highlights local inter-particle topological configurations. The advantages of 
the Voronoi PCF are demonstrated in several examples including crystalline, 
hyperuniform, and active systems showing clustering and giant 
number fluctuations. 
\end{abstract}

\maketitle  

\section{Introduction}
Many physical, chemical, and biological systems are studied as large  sets of 
point-like particles whose arrangement in space determines  many system 
properties.  Crystals have a relatively simple and elegant structure in which  
particle positions can be described with finite data (i.e., a periodically  
repeating unit cell).  However, in general, particles may be arranged 
in more  complex ways that cannot be described in such a simple manner 
\cite{torquato2002random}. The distribution of molecules in fluids 
\cite{terban2021structural},  celestial bodies in galaxies 
\cite{babu1996spatial}, trees in forests  \cite{Velazquez2016}, and animals 
moving in groups  \cite{vicsek1995novel} are but several examples. The 
appearance of discrete point patterns in multiple disciplines and  on various 
length scales highlights the need for general methods for  describing their 
structure, in both quantitative and qualitative terms 
\cite{diggle1976statistical,Ripley2005,Spodarev2013,morse2014geometric,
morse2016geometric,patania2017topological,carlsson2020topological,
skinner2021topological,skinner2023topological}.

The pair correlation function (\pcf) was introduced in the early  twentieth 
century as a method of describing structure in spatial point  patterns.  In 
particular, the \pcf details the distribution of distances  between particles 
within a system, and can therefore be used to infer,  among other features, 
system length scales. Another reason for the  tremendous success of the \pcf 
is its close relationship with X-ray  diffraction through the structure factor. 
These tools have facilitated  an accurate view of the atomic-level structure of 
countless materials  \cite{billinge2019rise}.

The \pcf, however, is limited in several respects.  Most importantly, it  is 
sensitive to small random perturbations, such as those caused by  
temperature or measurement errors, which do not typically affect structural 
classification \cite{kittel1976introduction}.  Post-processing  is subsequently 
necessary to distinguish artifacts associated with  noise from indicators of 
structurally significant features.  At the same time, it is rather insensitive to 
changes in a small fraction  of particles such as defects or local symmetries, 
that may be averaged  out globally, even though such details can have a 
significant effect on  macroscopic properties \cite{torquato2002random}. For 
these reasons, despite its great  success in providing deep understanding of 
crystalline and other  ordered systems, its utility in studying disordered ones 
has been  comparatively limited \cite{kardarfields, billinge2019rise}. 

In this paper we introduce a discrete version of the \pcf built on Voronoi 
topology \cite{lazar2015topological,leipold2016statistical}, and which we thus 
name the \textit{Voronoi pair correlation function} (\vpcf). It is complementary 
to the classical \pcf  in several respects: i.~It is naturally invariant under rescaling of 
length, as well as  translations and rotations. ii.~It is inherently robust to 
random perturbations and measurement errors \footnote{It is known that the 
Delaunay triangulation is almost surely unique and robust to infinitesimal 
perturbations. Degenerate cases, for example the square lattice, can be 
defined as the singular limit of vanishing Gaussian perturbations.}. iii.~It is 
sensitive to local order and symmetries. As  demonstrated below, the \vpcf 
proves useful in studying both ordered and  disordered systems.

Recall the definition of the two-point correlation function, as the two-
dimensional marginal distribution, $p^{(2)}$, normalized by the average
particle density $\rho$ \cite{kardarfields}. In many systems, $p^{(2)}$ only 
depends on distances between particles. One then defines the radial 
distribution function as
\begin{equation}   
g(r) = \frac{1}{\rho a(r)} p^{(2)} (|{\bf r}_1-{\bf r}_2|=r), 
\end{equation} 
where $a(r)$ is the perimeter or surface area of the $d$-dimensional spherical 
shell; $a(r)=2 \pi r$ in two dimensions and $a(r)=4 \pi r^2$ in three dimensions.  
Intuitively, if the probabilities of finding two particles at locations a distance 
$r_0$ apart are independent, then $g(r_0)=1$. For this reason, in systems 
with a finite correlation length, $g(r \to \infty)=1$. In particular, for an ideal gas 
with any density,  $g(r) \equiv 1$. In ideal periodic systems, the set of 
interparticle distances is necessarily discrete.  In more general systems, such 
as those described by thermodynamic ensembles, this set is typically 
continuous. Thus, $g(r)$ can distinguish between different ordered spatial 
configurations as well as some disordered ones 
\cite{Torquato2018,kardarfields}.

\overfullrule=0pt
\setlength{\fboxsep}{0pt}
\setlength{\tabcolsep}{0.2em}
\begin{figure*}
\begin{center}
\fbox{\begin{overpic}[trim={18.5mm 17.25mm 18.5mm 17.25mm},clip,height=0.397\columnwidth]{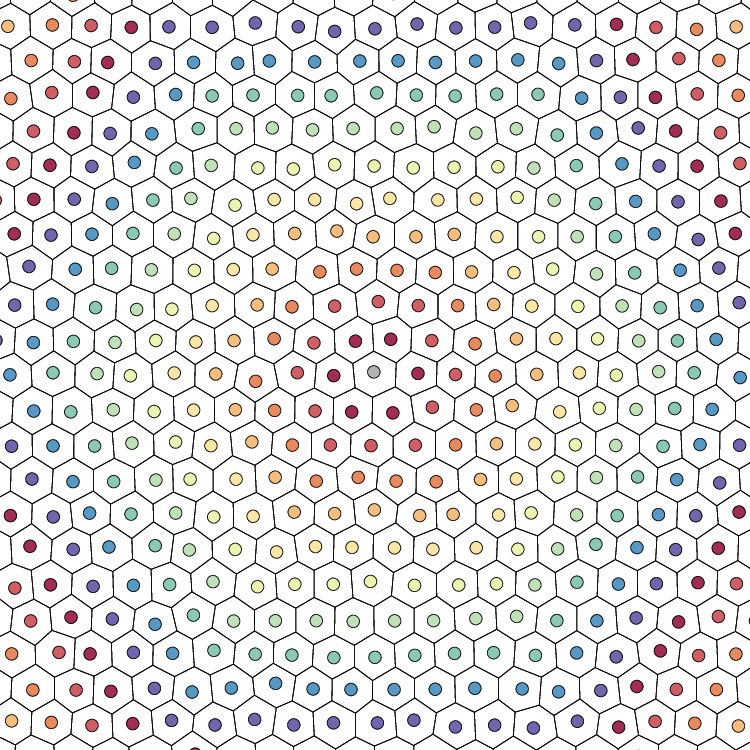}\put (-2.,90.) {\setlength{\fboxsep}{1pt} {\fcolorbox{black}{white}{(a)}}}  \end{overpic}}\hfill
\fbox{\begin{overpic}[trim={18.5mm 17.25mm 18.5mm 17.25mm},clip,height=0.397\columnwidth]{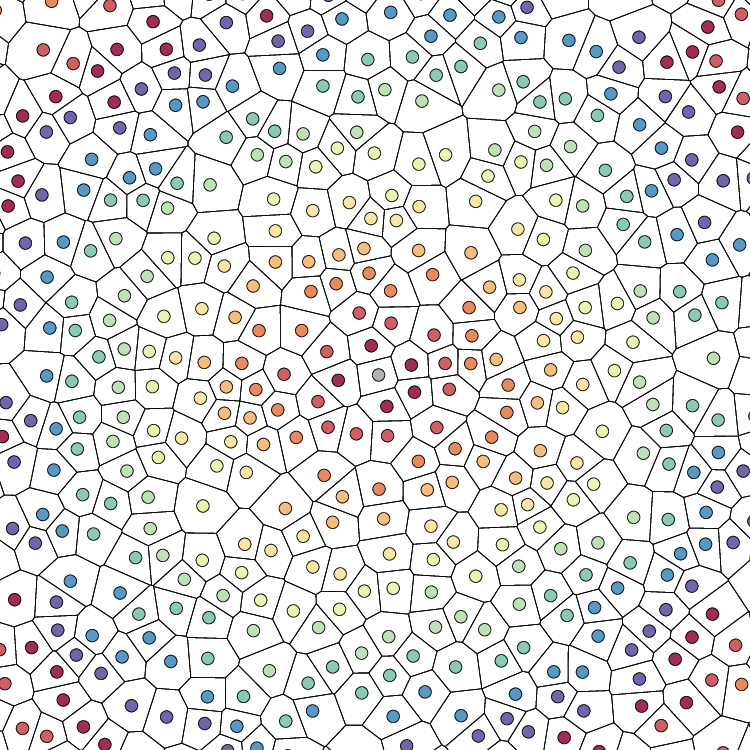}\put (-2.,90.) {\setlength{\fboxsep}{1pt} {\fcolorbox{black}{white}{(b)}}}  \end{overpic}}\hfill
\fbox{\begin{overpic}[trim={18.5mm 17.25mm 18.5mm 17.25mm},clip,height=0.397\columnwidth]{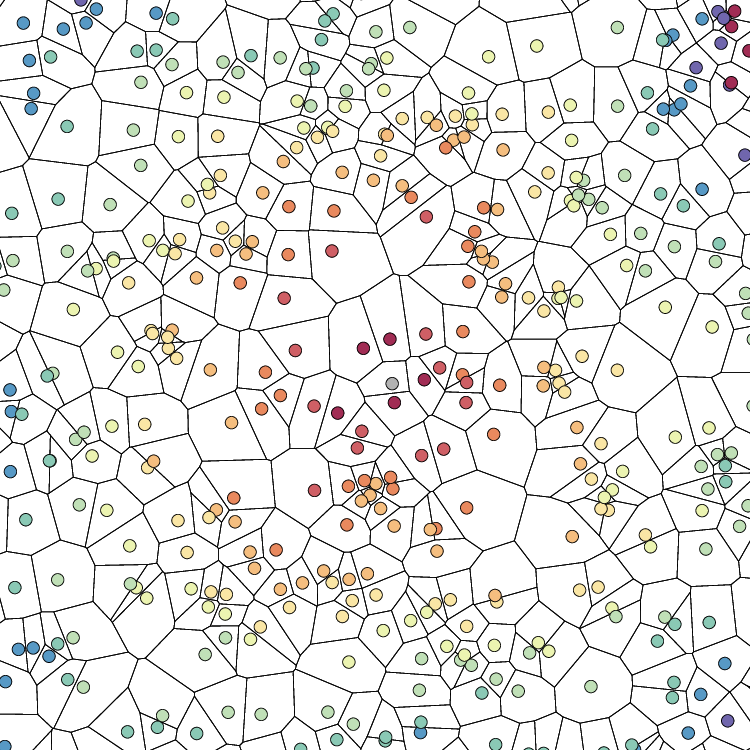}\put (-2.,90.) {\setlength{\fboxsep}{1pt} {\fcolorbox{black}{white}{(c)}}}  \end{overpic}}\hfill
\fbox{\begin{overpic}[trim={18.5mm 17.25mm 18.5mm 17.25mm},clip,height=0.397\columnwidth]{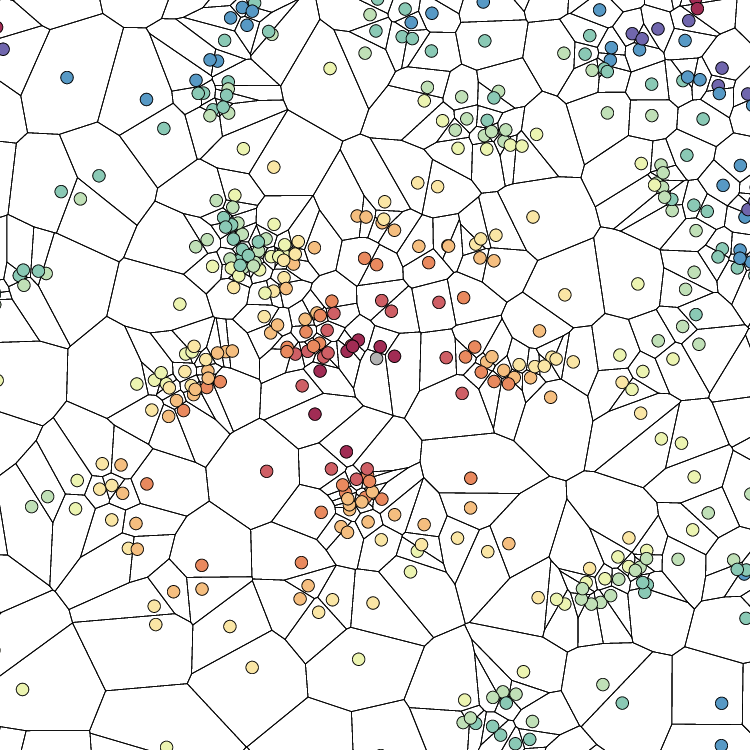}\put (-2.,90.)  {\setlength{\fboxsep}{1pt} {\fcolorbox{black}{white}{(d)}}}  \end{overpic}}\hfill
\fbox{\begin{overpic}[trim={18.5mm 17.25mm 18.5mm 17.25mm},clip,height=0.397\columnwidth]{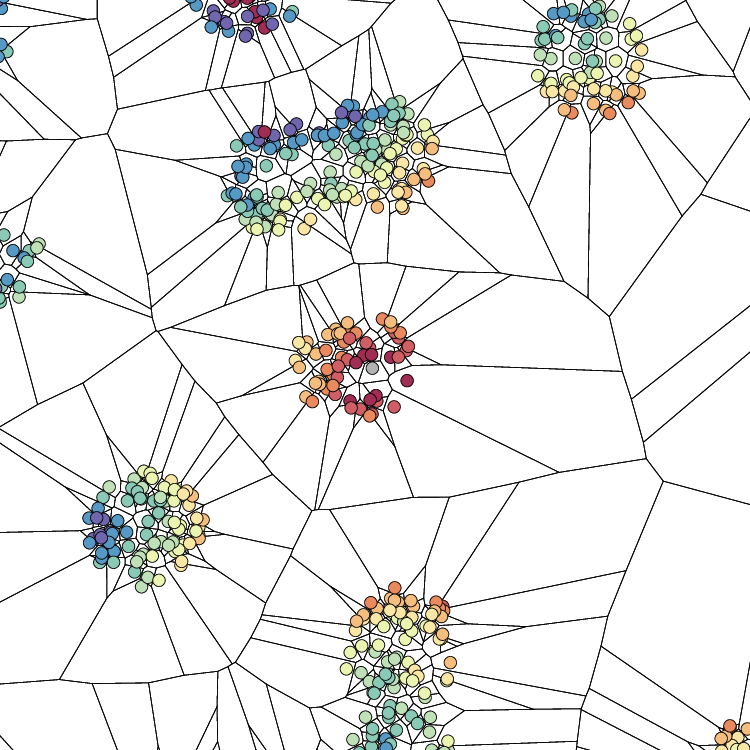}  \put (-2.,90.)  {\setlength{\fboxsep}{1pt} {\fcolorbox{black}{white}{(e)}}}\end{overpic}} \hfill 
\includegraphics[width=\linewidth, height=0.397\columnwidth, keepaspectratio]{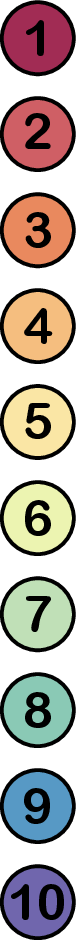}
\caption{A central particle and its Voronoi neighbors, colored by Voronoi 
distances from the central particle: 
(a) a low temperature, defect-free hexagonal crystal constructed using molecular dynamics and a Lennard-Jones potential, 
(b) a random organization example of a disordered hyperuniform system, 
(c) an ideal gas, 
(d) a Vicsek model of collective motion, and 
(e) a clustered Poisson process. A modified 
version of \textit{VoroTop} \cite{vorotop} has been used to illustrate the systems.
\label{vcells}}
\end{center}
\end{figure*}

\overfullrule=0pt
\setlength{\fboxsep}{0pt}
\setlength{\tabcolsep}{0.2em}
\begin{figure*}
\begin{center}
\fbox{\begin{overpic}[trim={10mm 9mm 9mm 9mm},clip,height=0.39\columnwidth]{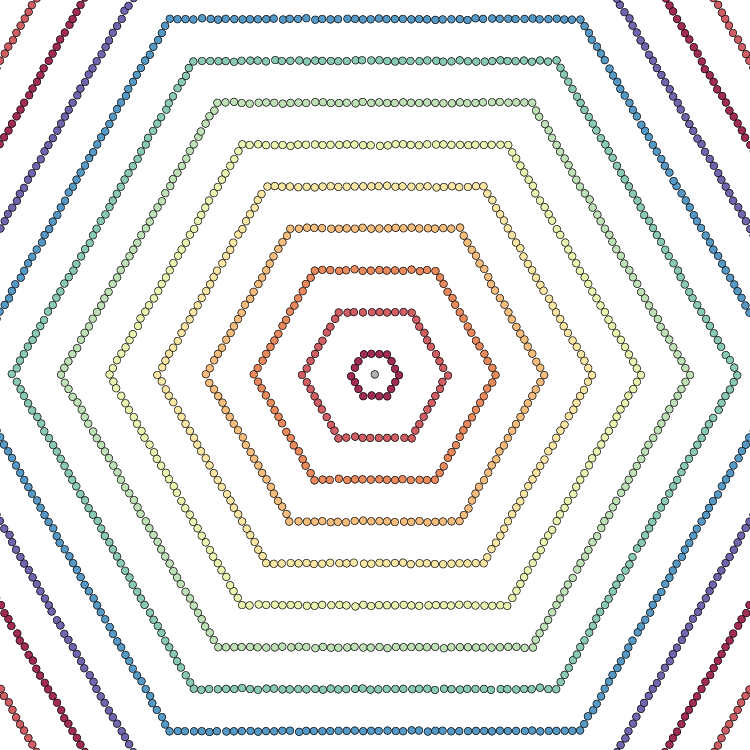}\put (-2.,90.) {\setlength{\fboxsep}{1pt} {\fcolorbox{black}{white}{(a)}}}  \end{overpic}}\hfill
\fbox{\begin{overpic}[trim={10mm 9mm 9mm 9mm},clip,height=0.39\columnwidth]{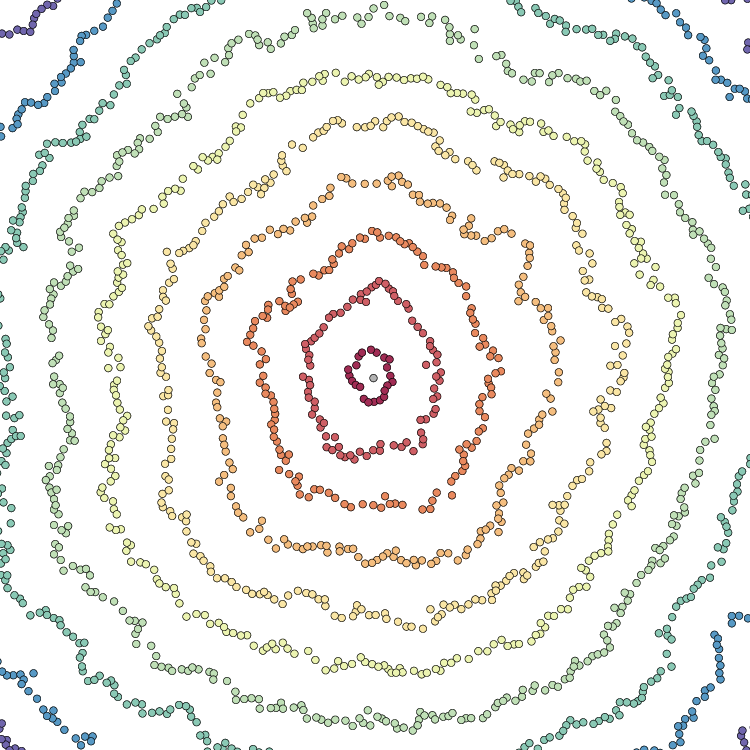}\put (-2.,90.) {\setlength{\fboxsep}{1pt} {\fcolorbox{black}{white}{(b)}}}  \end{overpic}}\hfill
\fbox{\begin{overpic}[trim={10mm 9mm 9mm 9mm},clip,height=0.39\columnwidth]{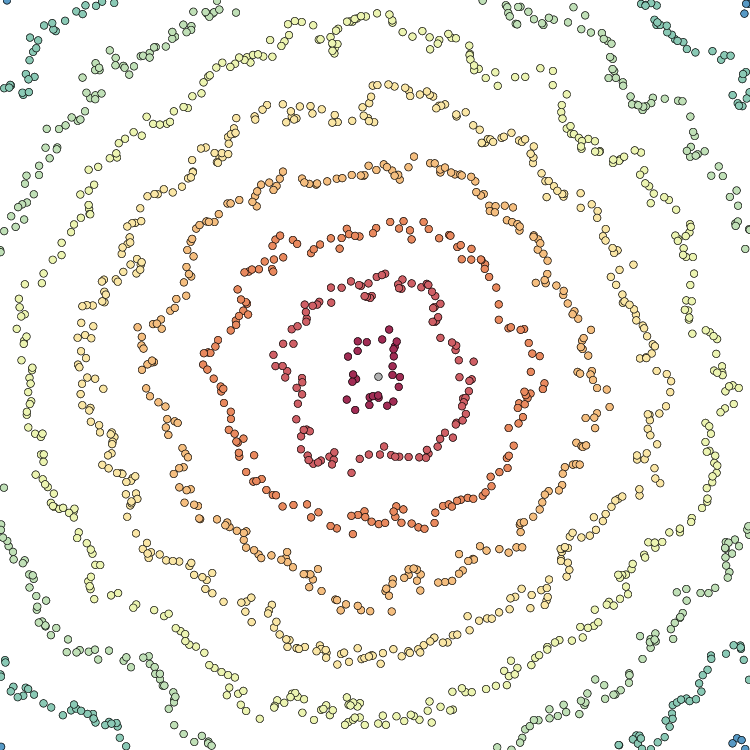}\put (-2.,90.) {\setlength{\fboxsep}{1pt} {\fcolorbox{black}{white}{(c)}}}  \end{overpic}}\hfill
\fbox{\begin{overpic}[trim={10mm 9mm 9mm 9mm},clip,height=0.39\columnwidth]{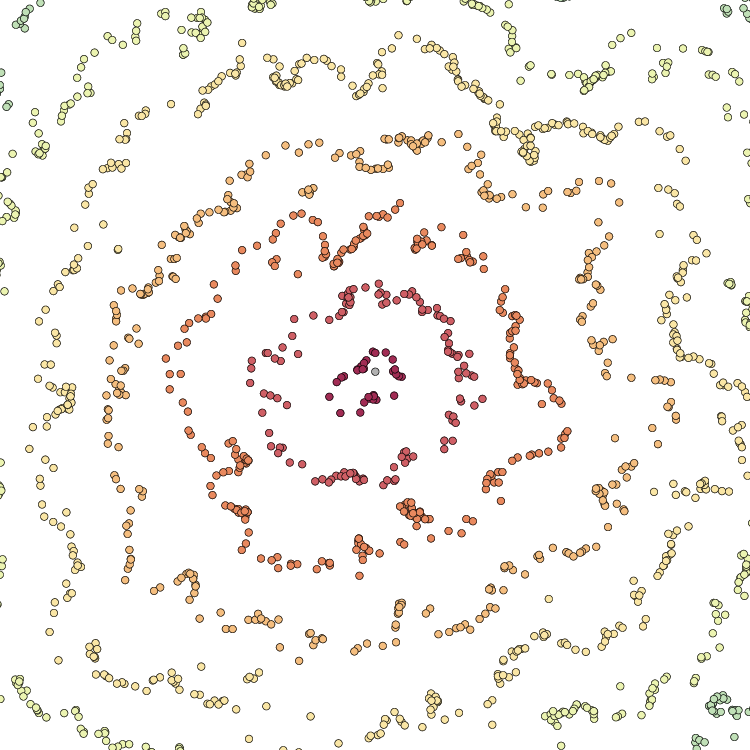}\put (-2.,90.)  {\setlength{\fboxsep}{1pt} {\fcolorbox{black}{white}{(d)}}}  \end{overpic}}\hfill
\fbox{\begin{overpic}[trim={10mm 9mm 9mm 9mm},clip,height=0.39\columnwidth]{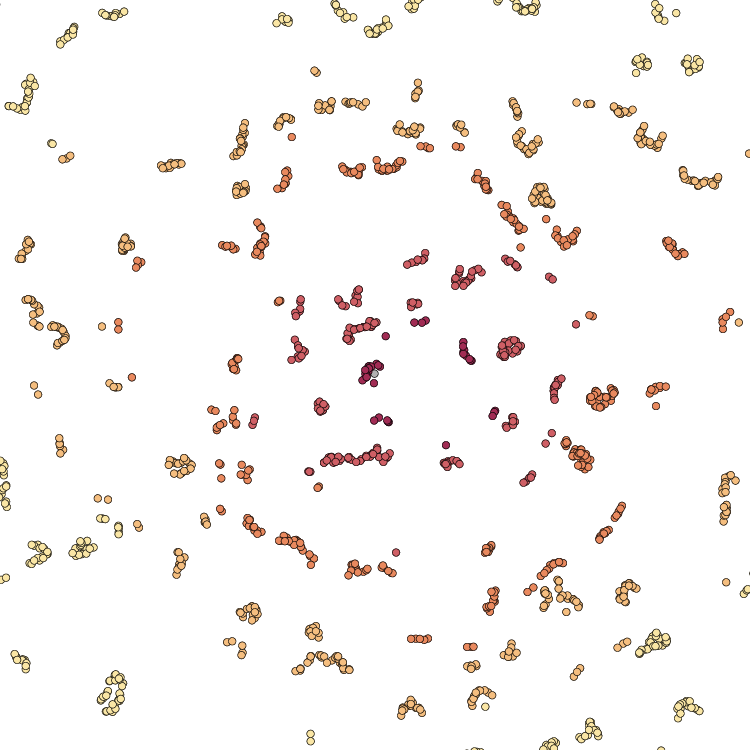}  \put (-2.,90.)  {\setlength{\fboxsep}{1pt} {\fcolorbox{black}{white}{(e)}}}\end{overpic}} \hfill 
\includegraphics[width=\linewidth, height=0.39\columnwidth, keepaspectratio]{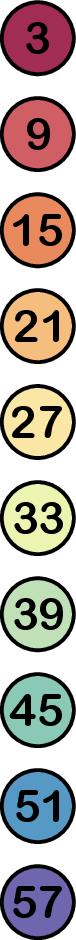}
\caption{A central particle and $k$-shells of its Voronoi neighbors, colored by 
Voronoi distances from the central particle: 
(a) a low temperature, defect-free hexagonal crystal constructed using molecular dynamics and a Lennard-Jones potential, 
(b) a random organization example of a disordered hyperuniform system, 
(c) an ideal gas, 
(d) a Vicsek model of collective motion, and 
(e) a clustered Poisson process. 
Quantitative information about these shells captures information about structural order.
\label{vcellsZoomOut}}
\end{center}
\end{figure*}

\section{A topological \\pair correlation function} 
To define a topological version of the \pcf, we first consider the Voronoi 
tessellation, a subdivision of a system into regions, called \textit{Voronoi 
cells}, that are closer to one particle than to any other 
\cite{voronoi1908nouvelles,lazar2022voronoi}.  Formally, given a discrete set 
of points (particle positions) $\{s_1, s_2,\ldots\} \subset \mathbb{R}^d$, the 
Voronoi cell of particle $i$ is the set
\begin{equation} 
V(s_i) = \{x \in \mathbb{R}^d \, | \, d(x,s_i) \leq d(x, s_j)\},  
\label{eqn:voronoi} 
\end{equation} 
where $d(x,y)$ is the standard Euclidean metric. In two dimensions, Voronoi 
cells are convex polygons, while in three dimensions they are convex 
polyhedra. 

Voronoi tessellations induce a discrete metric on particle systems. We  define 
distances between particles in terms of shortest paths, where path  lengths 
are quantified by the number of Voronoi cells that need be traversed  in 
moving from one particle to another.  Figure \ref{vcells} illustrates several two-dimensional 
systems and their Voronoi tessellations; particles are  colored according to 
their Voronoi distances from a central reference particle.  Figure 
\ref{vcellsZoomOut} depicts zoom-outs of the same systems, but  showing 
only several \textit{shells} -- sets of particles located at a fixed  Voronoi 
distance from a central reference particle. 

To quantify the patterns observed in the figures above, we next count the 
number of neighbors at each fixed Voronoi distance from a reference particle, 
and average over all particles.  In a defect-free hexagonal crystal, for 
example, each particle has exactly $6k$ neighbors at a Voronoi distance $k$; 
see Fig.~\ref{vcells}(a). 
This is the case even when particles are displaced by small perturbations, such 
as those resulting from thermal vibrations.  
In more general systems, however, there can be 
variation in the number of neighbors at different Voronoi distances; 
see Figs.~\ref{vcells}(b-e).  We use $k$-neighbors to refer to neighbors 
located at a Voronoi distance $k$ away from a reference particle, and 
$p_i(k)$ to denote the number of $k$-neighbors of particle $i$.  We use 
$u(k)$ to denote the average number of $k$-neighbors of particles in a 
system containing $N$ particles:
\begin{equation} 
u(k) = \left< \frac{1}{N}\sum_i^N p_i(k) \right>, 
\end{equation} 
where $\left< \cdot \right>$ denotes ensemble averaging and may also 
include variation in $N$. 

One of the nice properties of the classical \pcf is its normalization, which provides an 
intuitive physical meaning. In contrast to the classical \pcf, however, which 
requires rescaling of lengths, the \vpcf is determined by combinatorial 
features of Voronoi tessellations, which are naturally invariant under 
rescaling, and therefore no renormalization of length is necessary.   
We still, however, need to normalize $u(k)$, which diverges as $k\to \infty$, 
complicating the  comparison of systems.  Accordingly, and in analogy to the 
classical \pcf, we normalize $u(k)$  using data for an ideal gas, denoted 
$u_{\rm id} (k)$. We thus define the Voronoi pair correlation function $v(k)$ 
for a general particle system $S$ as
\begin{equation} 
v_S(k) = \frac{u_S(k)}{u_{\rm id}(k)}. 
\label{eq-normalization} 
\end{equation} 
In all systems considered here, $u(k)$ grows asymptotically linearly with $k$. 
This is expected under some reasonable assumptions for spatial systems such as a finite variance in the number of neighbors. The ratio between the 
asymptotic linear growth rate of $u_S(k)$ and that of $u_{\rm id}(k)$ provides 
the asymptotic value of $v_S(k)$ as $k\to\infty$. 
Determining the \vpcf only requires tools for the computation of Voronoi cells, 
or its dual Delaunay triangulation \cite{aurenhammer2013voronoi}, and 
shortest paths in networks. These tools are readily available in many software 
packages,  making the \vpcf simple to compute. Alternatively, \textit{VoroTop} 
\cite{vorotop} offers an efficient and parallelizable package of Voronoi 
topology computational tools, including the \vpcf.

\section{Examples}
\begin{figure}[b]
\begin{center}
\includegraphics[width=0.8\columnwidth]{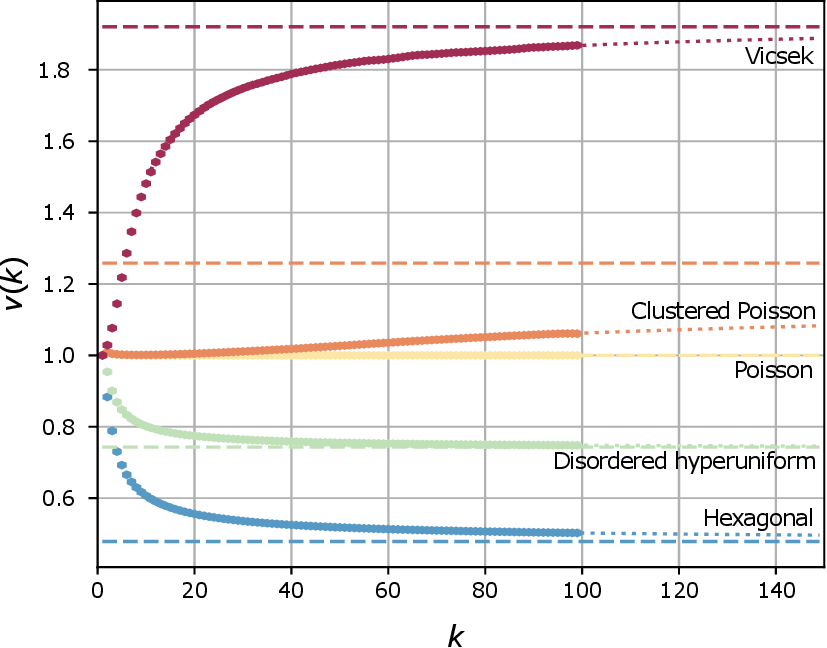}\vspace{-2mm}
\caption{The Voronoi pair correlation function $v(k)$ for the systems depicted 
in Fig.~\ref{vcells} with roughly 100,000 particles each. Shown are the data 
(solid points), a least-squares fit to Eq.~\ref{theeq} (dotted line), and the 
asymptotic values $v(\infty)$ (dashed line).\vspace{-5mm}}
\label{Growth}
\end{center}
\end{figure}

We now demonstrate the utility of the \vpcf by considering several 
two-dimensional systems illustrated in Figs.~\ref{vcells} and \ref{vcellsZoomOut}: 
a low-temperature, defect-free hexagonal crystal, 
a random organization example of a disordered hyperuniform system,
an ideal gas, 
a snapshot of a Vicsek simulation \footnote{Simulations follow 
\cite{vicsek1995novel} with $10^6$ particles, unit density, and uniform noise 
in $[-0.6\pi,0.6\pi]$; simulations were run for $10^5$ time steps.}, and a 
Poisson cluster process; Fig.~\ref{Growth} plots $v(k)$ for each.

\subsection{Poisson point process} 
A Poisson point process is a mathematical model of an ideal gas, in which 
points are independently and uniformly distributed in a fixed region 
\cite{lazar2013statistical}; see Fig.~\ref{vcells}(c). Although the full 
distribution of first nearest neighbors is known analytically 
\cite{calka2003explicit}, even the average number of $k$-neighbors for 
general $k$ is not.  

As mentioned earlier, we expect the average number of $k$-neighbors to 
grow asymptotically linearly with $k$, analogous to the linear growth of the 
perimeter of a circle with its radius.  Simulation data 
suggest that the average number of $k$-neighbors in equilibrium two-dimensional 
systems is approximated by a function of the form
\begin{equation}
u(k) = c_0 + c_1k + c_2k^{\frac{1}{2}} + c_3k^{\frac{1}{4}},
\label{theeq}
\end{equation}
ignoring lower-order terms, and with constants $c_i$ that are system specific.  
For the ideal gas, a least-squares fit of $u(k)$ constrained by $u(1)=6$ yields  
$c_0=18.77$, $c_1=12.58$, $c_2=-0.49$, and $c_3=-24.9$;
additional details can be found in the appendix.
The constant 
$c_1=12.58$ is noteworthy, since it is the linear growth rate of $k$-neighbors 
in the ideal gas, and will determine the asymptotic behavior of $v(k)$ for 
general systems.  

\subsection{Hexagonal lattice} The hexagonal crystal, illustrated in 
Fig.~\ref{vcells}(a), is among the simplest and most studied two-dimensional 
physical systems. As mentioned above, the number of $k$-neighbors of each 
particle is exactly $6k$, and so $u(k)=6k$. Dividing this by the linear term 
obtained for the ideal gas, $c_1=12.58$, we have $v(k\to \infty) = 0.477$.  
Figure \ref{Growth} shows that $v(k)$ for the hexagonal crystal is smallest 
among all systems considered.

\subsection{Poisson cluster process} 
A hallmark of active systems is giant number fluctuations 
\cite{ramaswamy2017active}. In these systems, activity drives clustering 
\cite{peruani2006nonequilibrium}, phase separation \cite{cates2015motility}, 
or arrested phase separation \cite{worlitzer2021motility}.  Activity and 
out-of-equilibrium statistics are thus manifested in spatial configurations. The 
effectiveness of the \vpcf in detecting such patterns is illustrated by 
consideration of a Mat\'{e}rn cluster point process, an example toy model of 
disordered systems that exhibits clustering \cite{Spodarev2013}. In this 
model, a set of ``parent'' points is generated through a homogeneous Poisson 
point process with intensity, or average density of points per unit area, $\phi$.     
Next, around each parent point is placed a disk of radius $r$ in which 
``children'' points are generated through independent homogeneous Poisson 
point processes with intensity $\lambda$. The result is a point process with 
intensity $\pi r^2\phi\lambda$, and in which points are grouped into clusters, 
possibly overlapping, with an average number of points $\pi r^2\lambda$ in 
each.  Figure \ref{vcells}(d) shows an example system with $\phi=1$, $r=0.2$, 
and $\lambda=500$, and Fig.~\ref{Growth} shows its \vpcf. In this case we 
find that $v(k \to \infty) = 1.26$. 

\begin{figure}
\begin{center}
\begin{overpic}
[width=1.\columnwidth]{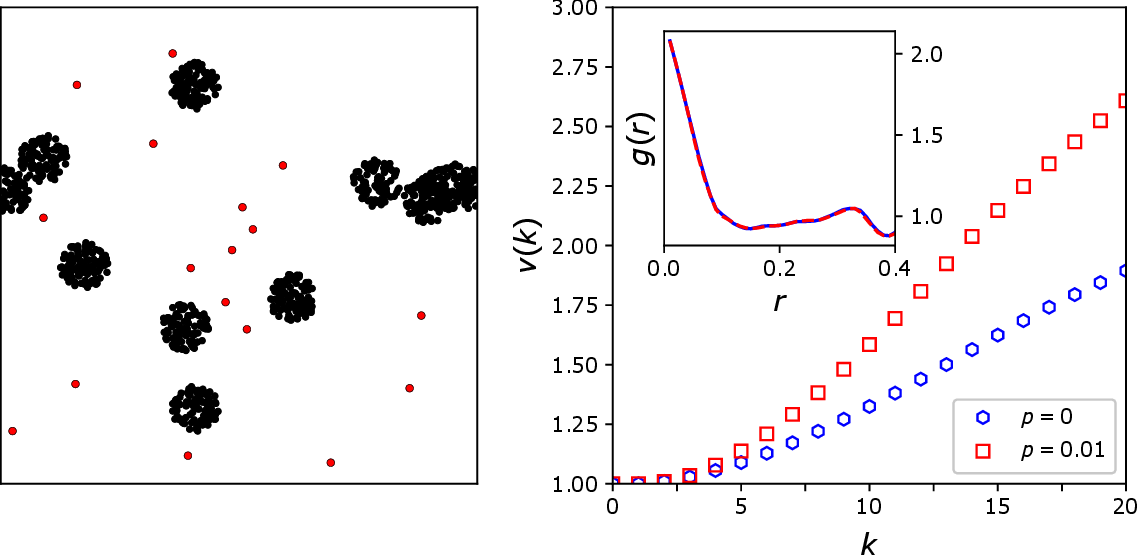}
\put (0.,44.5) {\setlength{\fboxsep}{1pt} {\fcolorbox{black}{white}{\footnotesize (a)}}}  
\put (91.,44.5) {\setlength{\fboxsep}{1pt} {\fcolorbox{black}{white}{\footnotesize (b)}}}  
\end{overpic}
\caption{(a) Poisson cluster process with $\phi=10$, $r=0.05$, and 
$\lambda=100$, and in which a fraction $p=0.01$ of children points are 
labeled as defects (red) and randomly distributed in space instead of in disks 
centered at parent points. (b) The Voronoi pair correlation function $v(k)$ for 
the original Mat\'{e}rn point process and the version with defects. Inset: the 
classical pair correlation functions of the two systems are indistinguishable.}
\label{Clustered}
\end{center}
\end{figure}

To highlight the sensitivity of the \vpcf to subtle differences in structure, we 
next consider a modified version of the Mat\'{e}rn cluster point process that 
includes ``defects''.  Specifically, while the majority of children points are 
located in clusters as described above, a fraction of them, $0<p<1$, are 
instead distributed randomly in the region.  These defect particles can be 
thought of as being generated by a Poisson point process with intensity 
$p\pi r^2\phi\lambda$; see Fig.~\ref{Clustered}(a) for an example when 
$p=0.01$.       

When $p$ is small, the vast majority of particles belong to densely populated 
disk-shaped clusters, and the presence of these isolated defects cannot be 
detected by the classical \pcf.  In contrast, these defects significantly alter the 
Voronoi tessellation and the resulting \vpcf. Figure \ref{Clustered}(b) 
illustrates the noticeable difference between the \vpcf of the original system 
and that with defects, demonstrating that the \vpcf can detect structural 
features that are ``hidden'' to the \pcf.  This might be of interest in analyzing 
networks in which hubs in otherwise unpopulated areas might significantly 
impact network structure by reducing path distances between points through 
local rewiring. 

\subsection{Disordered hyperuniformity} 
Hyperuniform structures are point patterns in which long-range spatial 
fluctuations are suppressed, as compared to an ideal gas 
\cite{Hexner2015,Hexner2017,Torquato2018}. More formally, a system is said 
to be \textit{hyperuniform} if its structure factor $S(q)$ vanishes for $q\to0$. 
Recently, it has been shown that hyperuniformity can be detected by 
considering the entropy of the many-body distribution of positions 
\cite{Ariel2020}. 

Examples of hyperuniform systems include crystals and quasicrystals. More 
intriguingly, disordered systems can exhibit hyperuniformity too.  
\begin{figure}
\begin{center}
\begin{overpic}[width=1.0\columnwidth]{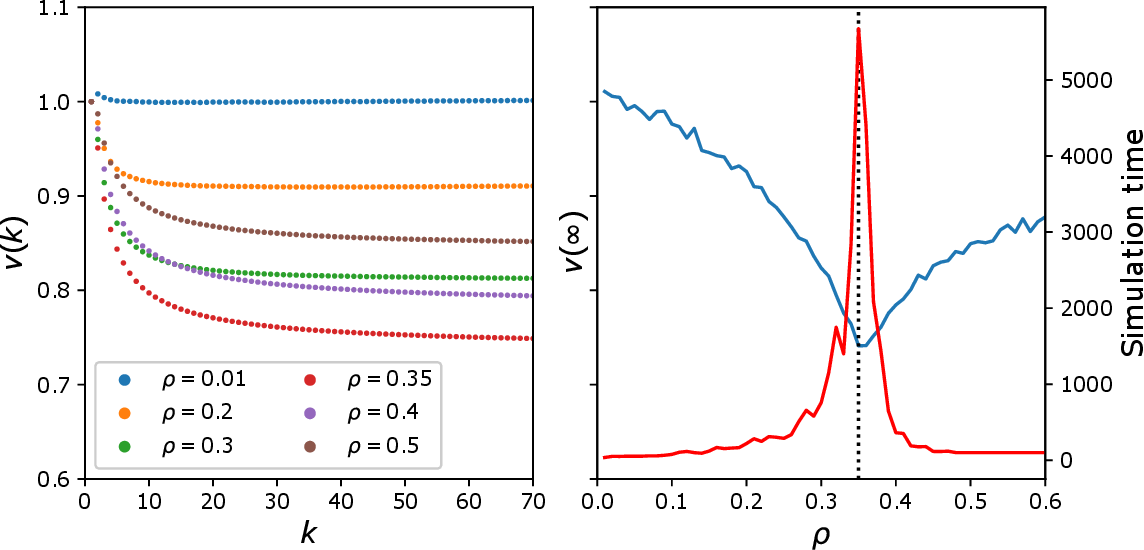}
\put (7.2,43.5) {\setlength{\fboxsep}{1pt} {\fcolorbox{black}{white}{\footnotesize (a)}}}  
\put (51.7,43.5) {\setlength{\fboxsep}{1pt} {\fcolorbox{black}{white}{\footnotesize (b)}}}  
\end{overpic}
\caption{(a) The Voronoi pair correlation function $v(k)$ for the random 
organization model at different densities $\rho$. (b) The simulation time (red 
line) to reach an absorbing state or a steady state, and the asymptotic value 
$v(\infty)$ while varying the densities (blue line); the scale is the same as in 
(a). The minimum of $v(\infty)$ at $\rho=0.35$ indicates the critical density for 
hyperuniformity (black dashed line).}
\label{Hyperuniform}
\end{center}
\end{figure}
One of the elegant models exhibiting disordered hyperuniformity is the 
\textit{random organization} model \cite{Hexner2015, Hexner2017}. In this 
model, $N$ disks of fixed diameter $r$ are randomly and independently 
placed in a domain $[0,L]^2$ with periodic boundaries. Overlapping disks are 
marked as active and are then displaced;  this procedure is repeated until 
either no disks overlap, in which case the system is said to have reached an 
absorbing state, or until the fraction of active disks is constant in time, 
in which case the system is said to have reached a steady state. The control 
parameter is the area fraction of disks, $\rho = N\pi r^2 / L^2$.  At low 
densities, an absorbing state is reached, while at high densities a steady 
state is reached. The transition occurs at a critical density $\rho_c$, which is 
smaller than the maximum packing fraction of disks in $[0,L]^2$.  
Interestingly, exactly at the transition, the system becomes hyperuniform, i.e., 
large scale density fluctuations are suppressed \cite{Hexner2015, 
Hexner2017,Torquato2018,Ariel2020}. 

Figure \ref{Hyperuniform}(a) shows the \vpcf for random organization models 
with different densities for 50,000 particles. At the critical density $\rho_c$, 
the \vpcf decreases fastest with increasing $k$. Indeed, the large distance 
limit $v(k \to \infty)$ exhibits a cusp with a minimum exactly at the critical density; see 
Fig.~\ref{Hyperuniform}(b). Intuitively, a hyperuniform system is similar to a 
crystal in terms of the number of Voronoi neighbors. As a result, the \vpcf is 
closer to that of a crystal than to that of a non-hyperuniform system; see 
Fig.~\ref{Growth}.

This example illustrates the strength of the \vpcf as a general-purpose tool 
for analyzing structure in particle systems.  In particular, the \vpcf can detect 
phase transitions which are characterized by structural changes.  Moreover, 
it is applied directly in real space rather 
than in Fourier space.  This is important, as it is often technically difficult to 
identify hyperuniformity from the classical \pcf or structure factor directly, as 
finite size effects often prevent the structure factor from vanishing at small 
wave numbers. In addition, sampling very short wave numbers may not be 
experimentally accessible \cite{ricouvier2017optimizing}.
However, note that $v(\infty)<1$ is not an indication of hyperuniformity.

\begin{figure*}
\begin{center}
\begin{overpic}
[trim={2.5mm 0.5mm 2.5mm 1mm},clip,width=0.69\columnwidth]{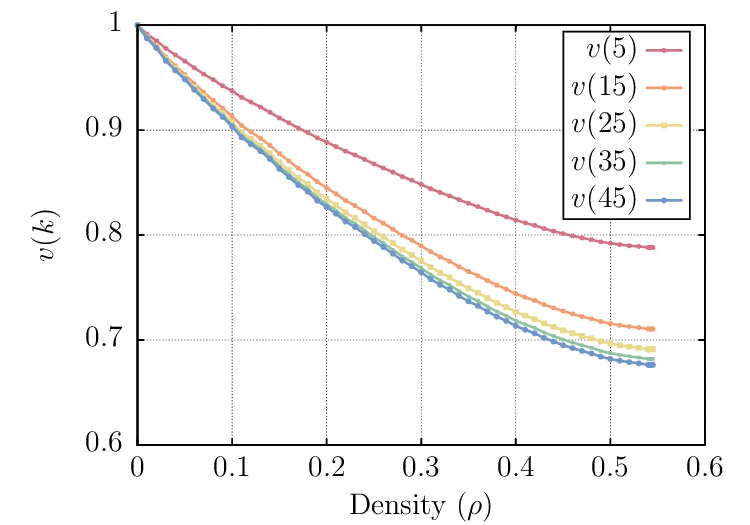}
\put (30,63) {\setlength{\fboxsep}{1pt} {\fcolorbox{black}{white}{\footnotesize (a)}}}  
\end{overpic}\hfill
\begin{overpic}
[trim={2.5mm 0.5mm 2.5mm 1mm},clip,width=0.69\columnwidth]{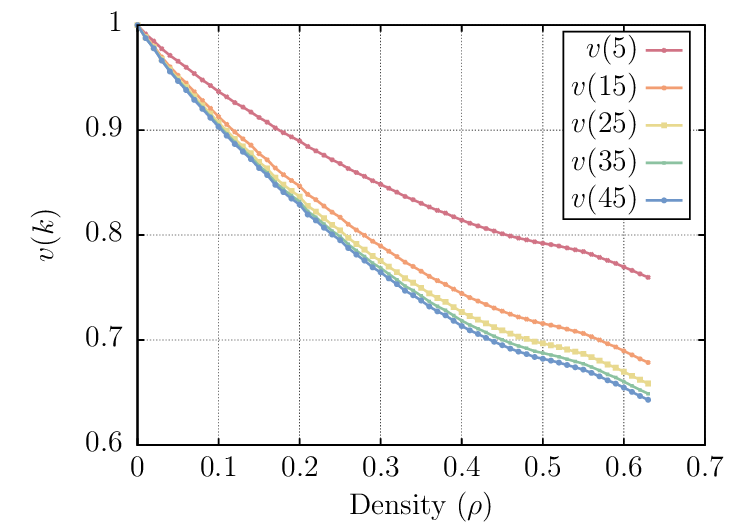} 
\put (30,63) {\setlength{\fboxsep}{1pt} {\fcolorbox{black}{white}{\footnotesize (b)}}}  
\end{overpic}\hfill
\begin{overpic}
[trim={2.5mm 0.5mm 2.5mm 1mm},clip,width=0.69\columnwidth]{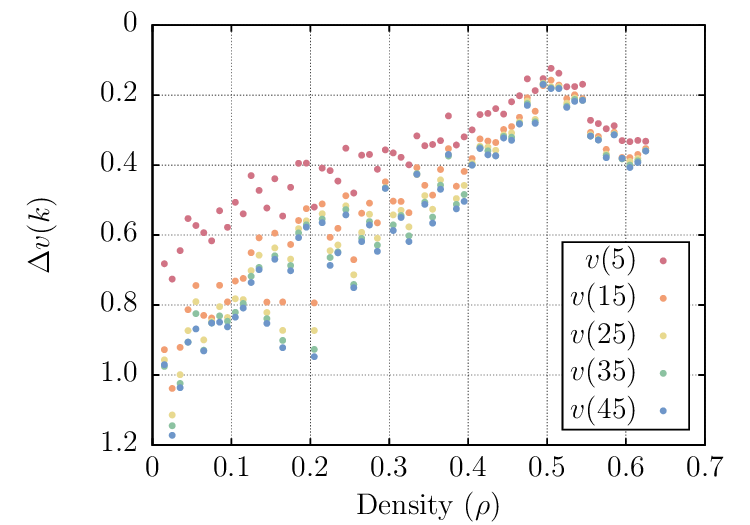}
\put (30,63) {\setlength{\fboxsep}{1pt} {\fcolorbox{black}{white}{\footnotesize (c)}}}  
\end{overpic}
\caption{The \vpcf for the (a) standard and (b) diffusion-modified RSA systems. 
A phase transition can be observed in the diffusion-modified system near 
$\rho=0.51$.  (c) This phase transition can be observed more clearly by 
considering the first finite difference of the \vpcf. 
\label{rsa}}
\end{center}
\end{figure*}

\subsection{Random sequential adsorption}
\label{appendix:rsa}
While this short paper cannot address the full breadth of hyperuniform systems \cite{torquato2021local},
we briefly consider a second example: random sequential adsorption (RSA) and 
a diffusion-modified version of it;
these are widely-studied models of adsorption and surface attachment 
\cite{lavalle1999extended,torquato2006random}. 

The standard RSA model in two dimensions is defined as follows.
Consider a domain $[0,L]^2$ with periodic boundary conditions.
Particles are placed sequentially into the domain. After placing $n$ 
particles, the position of particle $n+1$ is drawn uniformly within the 
domain. If the new position overlaps with any of the previous $n$ 
particles, it is discarded and redrawn. The process stops when no 
available positions remain. It has been shown numerically that the 
average maximal surface area is about 0.547 \cite{lavalle1999extended,torquato2006random}. 

We also implemented a variation of the model with surface diffusion \cite{lavalle1999extended}. Every time a position is discarded, the particles with which the candidate position overlapped are displaced randomly.
Displacements in each coordinate are taken as independent Gaussians with zero mean and standard deviation of 0.01.  

Figures \ref{rsa}(a) and (b) illustrate $v(k)$ for several $k$ over a range of densities in the standard and diffusion-modified systems.  Although the data are similar, the diffusion-modified system illustrates a (continuous) phase transition close to the critical density $\rho=0.51$.  Figure \ref{rsa}(c) shows a finite difference estimate of the derivative of $v(k)$ with respect to $\rho$. While $v(k)$ decreases with increasing density $\rho$, its second derivative changes sign at this critical point.

\subsection{Three dimensional examples}
\label{appendix:3d}
\begin{figure}[b]
\begin{center}
\vspace{2mm}
\includegraphics[width=0.84\columnwidth]{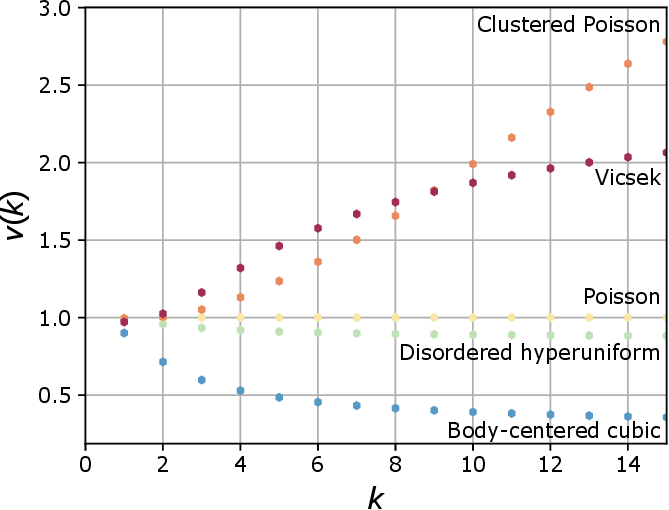}
\caption{The Voronoi pair correlation function for three-dimensional systems, each with roughly 500,000 particles. 
\vspace{-5mm}}
\label{Growth3D}
\end{center}
\end{figure}

Ideas developed in this paper immediately generalize to higher dimensions.  We briefly illustrate the application of the \vpcf to three-dimensional systems.  \textit{VoroTop} [23] was used to compute the average number of $k$-neighbors in a three-dimensional ideal gas and we then used those data to normalize the \vpcf of other particle systems described below and summarized in Fig.~\ref{Growth3D}.

\textit{The ideal gas.}
We constructed 50 independent copies, each with one million particles, and calculated the \vpcf in each of them. 
In two dimensions the number of $k$-neighbors grows linearly in $k$, as the perimeter of a circle is proportional to its radius.  The proportion constant quantifies the discrete nature of the conversion from a continuous perimeter to a discretized $k$-shell, and also the degree to which the $k$-shells are circular.  In three dimensions, the surface area of a sphere grows as the square of its radius, and so we expect that the number of $k$-neighbors, a discrete analog of surface area, will grow approximately linearly in $k^2$.  
We thus considered functions of the form:
\begin{equation}
u(k) = c_0 + c_1k + c_2k^2
\tag{6}
\end{equation}
to describe the number of $k$-neighbors in a three-dimensional ideal gas.  A least squares fit of the simulation data results in constants $c_0=381$, $c_1=-166$, and $c_2=42.7$.  
We note that while this form accurately fits data for large $k$, it is less accurate for small $k$.  For this reason we use actual numerical data to normalize $v(k)$ for other systems.

\textit{Perturbed crystals}. 
The three-dimensional \textit{body-centered cubic} crystal can be considered an analog of the two-dimensional hexagonal crystal in the sense that its Voronoi cells are topologically stable -- small perturbations of particle positions do not result in topological changes in the Voronoi cells.  In this system, each Voronoi cell is a truncated octahedron with eight hexagonal faces and six square ones.  
Figure \ref{Growth3D} shows that $v(k)$ is smallest for the body-centered cubic crystal among all three-dimensional systems considered. 

\textit{Poisson cluster process.} 
We constructed Poisson cluster processes in the same manner as in two dimensions.  In particular, we studied a system with $\phi=10$, $r=0.05$, and $\lambda=100$.
The apparent unbounded growth of $v(k)$ for this system might be attributed to the system sizes considered.  With the increase in dimension, the number of $k$-neighbors grow, and hence significantly larger systems must be considered to obtain accurate estimates of $v(k)$ for large $k$.  Meanwhile, the computational expense to compute them also increases.  

\textit{Vicsek model.}
We also looked at a quasi-three dimensional system derived from a Vicsek model.  We ran a classical Vicsek model \cite{vicsek1995novel} with $10^6$ particles, unit density, and uniform noise in $[-0.6\pi,0.6\pi]$; simulations were run for $10^5$ time steps.  At the end of the simulations, we rescaled the particle directions, which normally are measured in angles in $[0,2\pi]$ with periodic boundary conditions to lie in $[0,1]$ instead. Thus, we obtain a quasi-three dimensional representation of the data.

\begin{figure}
\begin{center}
\begin{overpic}
[width=1.\columnwidth]{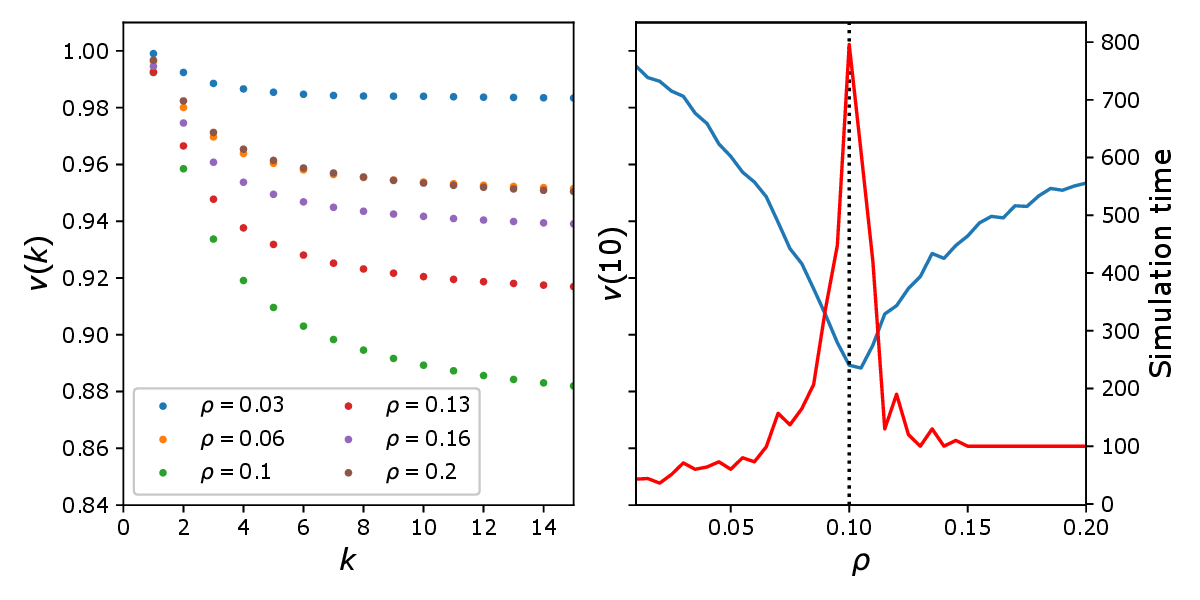}
\put (40.2,44.5) {\setlength{\fboxsep}{1pt} {\fcolorbox{black}{white}{\footnotesize (a)}}}  
\put (82.75,44.5) {\setlength{\fboxsep}{1pt} {\fcolorbox{black}{white}{\footnotesize (b)}}}  
\end{overpic}
\caption{(a) The Voronoi pair correlation function $v(k)$ for the random 
organization model in three dimensions at different densities $\rho$. (b) The simulation time (red line) to reach an absorbing state or a steady state, and the asymptotic value 
$v(\infty)$ while varying the densities (blue line); the scale is the same as in 
(a). The minimum of $v(10)$ at $\rho=0.10$ indicates the critical density for 
hyperuniformity (black dashed line).}
\label{Hyperuniform3D}
\end{center}
\end{figure}

\textit{Disordered hyperuniform.}
We constructed three-dimensional analogs of the two-dimensional random organization model described above.
Figure \ref{Hyperuniform3D} illustrates the same critical behavior in three dimensions observed in two dimensions.  In particular, when $\rho$ is below a critical threshold, in this case roughly $0.1$, the \vpcf decreases with increasing density $\rho$, after which it increases.  This criticality is mirrored in the simulation time necessary to reach an absorbing or steady state.

\section{Discussion} 
Recent decades have seen the rapid development and application of 
topological methods in analyzing scientific data 
\cite{patania2017topological,carlsson2020topological}, contributing to 
progress in many problems in the physical and biological sciences 
\cite{skinner2021topological,skinner2023topological}. Much of the success of 
such methods results from their segmentation of data in high-dimensional 
configuration spaces associated with the problem, instead of in 
lower-dimensional images of those spaces obtained through continuous functions 
\cite{landweber2016fiber}.  

The discrete Voronoi pair correlation function provides a natural topological 
version of the classical pair correlation function. The \vpcf succinctly captures 
structural information about particle systems in a manner that is naturally 
scale invariant, and which is also stable with respect to small perturbations, 
such as those associated with temperature or measurement error. At the 
same time it is able to detect structural features ``hidden'' from classical, 
purely geometric, methods. This might be useful, for example, in studying the 
glass transition, which is commonly believed to occur without or with only 
minimal structural changes \cite{janssen2018mode}, and for which the 
classical \pcf is ineffective. 

The present paper complements recent advances in the application of
Voronoi analysis to the study of ordered and disordered particle systems. 
Morse and Corwin considered the average number of nearest Voronoi 
neighbors, as well as geometric features of the Voronoi cells, to classify the 
jamming transition \cite{morse2014geometric}.  
Klatt and Torquato looked at volumes, surface areas and mean widths of 
Voronoi cells to characterize the structure of maximally random jammed 
sphere packings \cite{klatt2014characterization}.  They further studied 
correlations of these geometric features among neighboring Voronoi cells 
to provide a more refined structural description of disordered particle 
systems \cite{klatt2014characterization}.
In contrast to these predominantly geometric approaches, Skinner and 
coauthors \cite{skinner2021topological,skinner2023topological} develop
a topological approach towards quantifying the similarity and difference 
between Delaunay triangulations obtained from disordered systems.  Their method was used to 
identify structural differences among several biological systems, highlighting the scientific diversity of applications that can benefit from topological methods.

We note that although $v(k)$ captures much structural information in a 
discrete sequence of numbers, important information is lost due to averaging 
data over all particles.  We can thus also consider higher-order moments of 
the distribution of $k$-neighbors.  Such information might distinguish distinct 
crystal structures that cannot be distinguished when considering $v(k)$ alone.  

Finally, we have focused most of our attention on two-dimensional systems largely for 
simplicity of presentation and for ease of illustration. 
We anticipate that 
higher dimensions will provide interesting challenges as well as opportunities 
that do not arise in two dimensions.  In particular, the shapes of $k$-shells 
are expected to have significantly richer features.

\textbf{Acknowledgements.} 
This work was supported by the Data Science Institute at Bar-Ilan University.   

\bibliographystyle{ieeetr}

\begin{figure}
\begin{center}
\includegraphics[width=0.85\columnwidth]{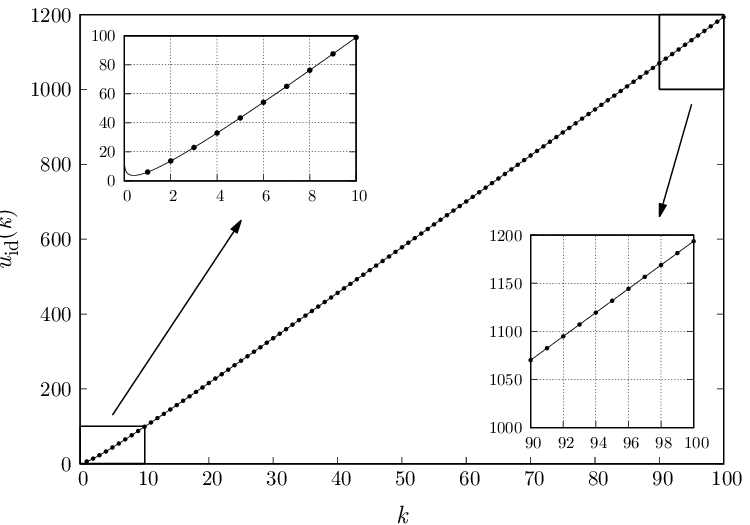}
\caption{Points illustrate the average number of $k$-neighbors of all particles in 100 samples of an ideal gas in two dimensions, each containing half a million points. The fitted curve is of the form given by Eq.~\ref{theeq}; error bars are too small to see.}
\label{uk}
\end{center}
\end{figure}

\appendix
\setcounter{secnumdepth}{0}

\section{Appendix: semi-analytic form of $u_{PV}(k)$}
\label{appendix:upv}

Simulation data suggest that the average number of $k$-neighbors in equilibrium 
two-dimensional systems is accurately approximated by a function of the form
\begin{equation}
u(k) = c_0 + c_1k + c_2k^{\frac{1}{2}} + c_3k^{\frac{1}{4}}.
\tag{5}
\end{equation}
In particular, the number of $k$-neighbors grows asymptotically linearly in $k$.  
This can be viewed as a discretized version of the linear relationship between the perimeter of a circle, discretized in $k$-shells, and its radius. 

To obtain accurate estimates of the parameters $c_i$ for the homogeneous Poisson 
point process, we constructed 100 systems, each containing half a million particles 
uniformly distributed in a square with periodic boundaries, 
and computed $u(k)$ in each of them for $1 \leq k \leq 100$.
Averaging the data and using a weighted least squares fit to Eq.~\ref{theeq}, constrained so that $u(1)=6$, yields $c_0=18.77$, $c_1=12.58$, $c_2=-0.49$, and $c_3=-24.9$, with the least uncertainty in $c_1$.   
This provides an excellent fit to the observed data for both large and small $k$; the root mean square deviation is $0.01$.

Figure \ref{uk} illustrates these data and a fit of $u(k)$ using the parameters $c_i$ as described.  As can be seen from the inset figures, Eq.~\ref{theeq} with the fitted parameters accurately approximates the observed data for both low and high values of $k$.

\end{document}